\documentclass[useAMS,usenatbib]{emulateapj}
\usepackage{graphicx}
\usepackage{multirow}
\usepackage{amsmath}
\usepackage{array}
\usepackage{times}
\usepackage{pifont}
\usepackage[colorlinks=false]{hyperref}
\slugcomment{Astrophysical Journal, in press (Accepted 2013 May).}

\shorttitle{Suzaku observation of ESO~565--G019}
\shortauthors{Gandhi et al.}

\def\c{{\em Chandra}}

\def\swift{{\em Swift}}

\def\astroh{{\sl Astro-H}}
\def\akari{{\sl Akari}}
\def\wise{{\sl WISE}}
\def\nustar{{\sl NuSTAR}}

\def\suzaku{{\em Suzaku}}

\def\iras{{\em IRAS}}
\def\herschel{{\em Herschel}}

\def\p{$\pm$}
\def\ltsim{\mathrel{\hbox{\rlap{\hbox{\lower4pt\hbox{$\sim$}}}\hbox{$<$}}}}
\def\gtsim{\mathrel{\hbox{\rlap{\hbox{\lower4pt\hbox{$\sim$}}}\hbox{$>$}}}}

\def\Msun{M$_{\odot}$}
\def\Lsun{L$_{\odot}$}
\def\micron{$\mu$m}
\def\araa{ARA\&A}
\def\aap{A\&A}
\def\nat{Nat}

\def\aaps{A\&AS}
\def\mnras{MNRAS}
\def\apj{ApJ}
\def\apjs{ApJS}
\def\aj{AJ}
\def\apjl{ApJL}

\def\pasj{PASJ}

\def\hi{H~{\sc i}}

\def\nh{$N_{\rm H}$}

\def\eso565{ESO~565--G019}
\def\xspec{{\sc xspec}}
\def\apec{{\sc apec}}
\def\pexrav{{\sc pexrav}}
\def\mytorus{{\sc mytorus}}

\begin{document}

\title{Reflection-dominated nuclear X-ray emission in the early-type galaxy ESO~565--G019}
\author{P. Gandhi\altaffilmark{1,2}, Y. Terashima\altaffilmark{3}, S. Yamada\altaffilmark{4}, R.F. Mushotzky\altaffilmark{5}, Y. Ueda\altaffilmark{6}, W.H. Baumgartner\altaffilmark{7}, D.M. Alexander\altaffilmark{2}, J. Malzac\altaffilmark{8,9}, K. Vaghmare\altaffilmark{10}, T. Takahashi\altaffilmark{1,11} and C. Done\altaffilmark{2}}
\altaffiltext{1}{Institute of Space and Astronautical Science (ISAS), Japan Aerospace Exploration Agency, 3-1-1 Yoshinodai, chuo-ku, Sagamihara, Kanagawa 252-5210, Japan}
\altaffiltext{2}{Department of Physics, Durham University, Durham DH1-3LE, UK}
\altaffiltext{3}{Department of Physics, Ehime University, 2-5, Bunkyo-cho, Matsuyama, Ehime 790-8577, Japan}
\altaffiltext{4}{Cosmic Radiation Laboratory, Institute of Physical and Chemical Research (RIKEN), Wako, Saitama, 351-0198, Japan}
\altaffiltext{5}{Department of Astronomy, University of Maryland College Park, College Park, MD 20742}
\altaffiltext{6}{Department of Astronomy, Kyoto University, Kitashirakawa-Oiwake-cho, Sakyo-ku, Kyoto 606-8502, Japan}
\altaffiltext{7}{NASA/Goddard Space Flight Center, Astrophysics Science Division, Greenbelt, MD 20771}
\altaffiltext{8}{Universit\'{e} de Toulouse, UPS-OMP, IRAP, Toulouse, France}
\altaffiltext{9}{CNRS, IRAP, 9 Av. colonel Roche, BP 44346, F-31028 Toulouse cedex 4, France}
\altaffiltext{10}{IUCAA, Post Bag 4, Ganeshkhind, Pune 411007, India}
\altaffiltext{11}{Department of Physics, University of Tokyo, 7-3-1 Hongo, Bunkyo, Tokyo, 113-0033, Japan}
\label{firstpage}
\begin{abstract}
We present the discovery of a reflection-dominated active galactic nucleus (AGN) in the early-type radio-quiet galaxy ESO~565--G019 with \suzaku\ and \swift/BAT. The source X-ray spectrum below 10 keV is characteristic of other Compton-thick (CT) AGN, clearly showing an inverted continuum and prodigious fluorescence iron emission above $\sim$3 keV. A Compton shoulder to the neutral Fe K$\alpha$ line also appears to be present. There is evidence for long-term hard X-ray flux variability which we associate with changes in the intrinsic AGN power-law. The increasing sensitivity of ongoing and new hard X-ray surveys means that more such reflection-dominated AGN ought to be uncovered in the near future. ESO~565--G019 is hosted in an early-type galaxy whose morphology has been variously classified as either type E or type S0. Only about 20 bona fide CT-AGN have been identified in the local universe so far, and all exist in host galaxies with late Hubble types (S0 or later). CT columns of nuclear obscuring gas are uncommon in early-type galaxies in the local universe, so confirmation of the exact morphological class of ESO~565--G019 is important. Infrared photometry also shows the presence of large quantities of cool dust in the host, indicative of significant ongoing star-formation. ESO~565--G019 may be the first identified local example of minor-merger driven CT-AGN growth in an early-type host, or may be the result of interaction with its neighboring galaxy ESO~565--G018 in a wide pair. 
\end{abstract}

\section{Introduction}

The spectrum of the cosmic hard X-ray background requires a substantial contribution by heavily obscured active galactic nuclei \citep[AGN; e.g. ][]{settiwoltjer89, fi, g03, gilli07, treister09, ballantyne11, brightmanueda12, akylas12}. Extensive searches have been carried out for the most obscured population of objects using a range of multi-wavelength selection techniques \citep[e.g., ][ and more]{bassani99, risaliti99, g04, guainazzi05_obscuration, martinezsansigre07, fiore08_comptonthick, goulding09, alexander11, severgnini12, georgantopoulos13}. %

Each technique suffers from some form of incompleteness. In particular, Compton-thick (CT) AGN with equivalent neutral Hydrogen column densities of obscuring gas \nh$\gtsim$1.2--1.5$\times$10$^{24}$ cm$^{-2}$ are notoriously difficult to identify, even in the local universe. Their X-ray spectra are expected to peak in the hard X-ray regime above 10 keV \citep[e.g. ][]{matt00}, but Compton-downscattering can significantly attenuate even hard X-rays. X-rays from nuclear stellar activity or from the AGN component scattered into the line-of-sight can then overwhelm obscured AGN signatures. 

Here, we present high quality \suzaku\ follow-up of a galaxy selected based upon its hard X-ray properties from the 70-month \swift\ Burst Alert Telescope (BAT) sky survey \citep{baumgartner13}. \eso565\ is generally classified as optical morphological type E\footnote{http://leda.univ-lyon1.fr/}, hosting an optically-classified Seyfert 2 \citep{veron06}. No pointed X-ray observations below 10 keV have been published. Its detection in the \swift/BAT survey at higher energies suggests that the source may have been previously overlooked. We find a reflection-dominated X-ray continuum and strong Fe fluorescence emission lines, and show that the AGN is likely to be hidden by CT columns of obscuring gas. CT-AGN activity in early-type galaxies is relatively uncommon in the local universe, and we discuss how \eso565\ fits into this context.  

Luminosities are based on a redshift $z$=0.017379 (72.4 Mpc) corrected to the reference frame of the cosmic microwave background for $H_0$=73 km~s$^{-1}$~Mpc$^{-1}$, $\Omega_{\rm M}$=0.27, $\Omega_{\Lambda}$=0.73.\footnote{http://ned.ipac.caltech.edu}

\section{Observations}

\subsection{\swift/BAT}

The BAT instrument on-board the \swift\ mission \citep{swift, bat} has been used to compile the 70-month hard X-ray catalog (\citealt{baumgartner13}; see also {\tt http://heasarc.gsfc.nasa.gov/docs/swift/} {\tt results/bs70mon}) including publicly-released spectra and diagonal response matrices, which we use here. The effective source exposure is 8.1 Ms. The source displays a relatively-flat power-law photon-index of $\Gamma$=1.37$_{-0.40}^{+0.39}$ and 14--195 keV flux of 2.1(\p0.6)$\times$$10^{-11}$ erg s$^{-1}$ cm$^{-2}$. The corresponding luminosity is 1.4$\times$10$^{43}$ erg~s$^{-1}$.

\subsection{\suzaku}

\eso565\ was observed with \suzaku\ \citep{suzaku} starting on 2012 May 20 for an on-source effective time of 78.9 ks; ObsID=707013010. Both instruments, the X-ray Imaging Spectrometer (XIS; \citealt{xis}) and the Hard X-ray Detector (HXD; \citealt{hxd, hxdinorbit}) were operated in standard mode with XIS-nominal pointing. 

Data reduction and processing were performed using the \suzaku\ software version 18, {\sc ftools} v6.12 \citep{ftools}. Standard recommended event selection criteria\footnote{http://heasarc.gsfc.nasa.gov/docs/suzaku/analysis/abc} were employed using {\sc xselect} v2.4b. The net XIS and HXD good times were 62.9 and 66.5 ks respectively. 

XIS (sensitive below 10~keV) counts were integrated within a circular aperture of radius 3\farcm 4, and background counts extracted from source-free regions. \eso565\ is by far the brightest source within the XIS field-of-view (fov). A polygonal background aperture was constructed to sample large parts of the background over the remaining fov, free from contamination by other possible faint sources. The latest version of the calibration database v20120902 was used for generating response matrices and auxiliary response files. 

For HXD, data from the PIN diodes below $\sim$60 keV were extracted using the {\sc ftool} {\tt hxdpinxbpi}, which returns the deadtime-corrected source spectrum and a background spectrum based upon the \lq tuned\rq\ background model provided by the \suzaku\ team. Below $\sim$30 keV, this model was found to severely overestimate the earth-occultation background data\footnote{selected using events with ELV$<$--5; effective exposure time 41.7 ks}, which were also extracted for comparison. This is a result of insufficient samples for good model reconstruction of the long-term background trend during mid-May 2012, and the model was discarded in favor of the earth-occultation data. The contribution of the cosmic X-ray background component was simulated in a standard manner\footnote{http://heasarc.gsfc.nasa.gov/docs/suzaku/analysis/abc} and added to the earth-occultation spectrum. For model-fitting using $\chi^2$ minimization, spectral counts were grouped to obtain a minimum significance of 4 per bin. At higher energies, the source is faint enough that data from the scintillator array (GSO) were ignored. 

\begin{figure*}
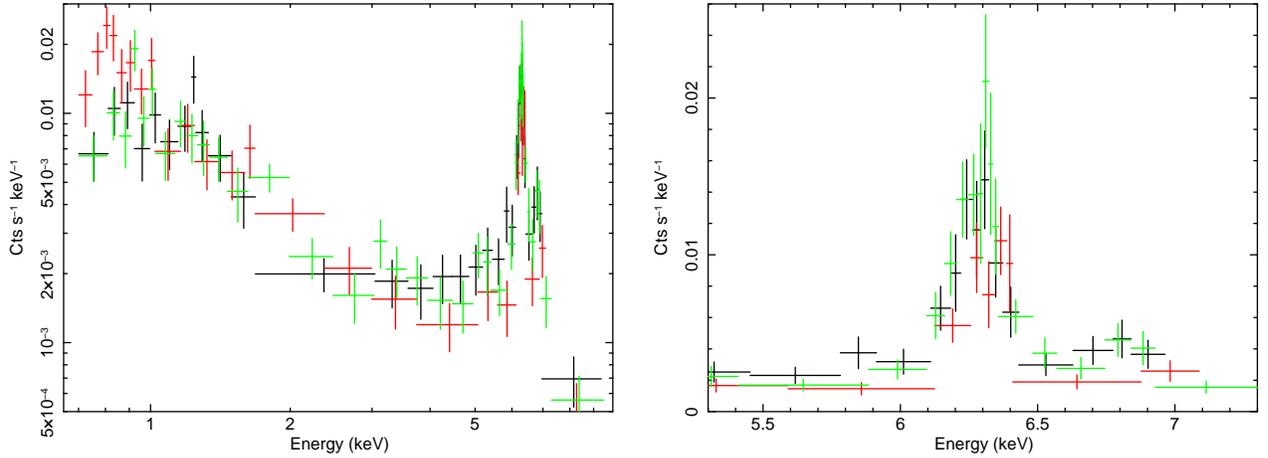

  \begin{center}
    \includegraphics[angle=270,width=8.5cm]{f1a.ps}
    \includegraphics[angle=270,width=8.5cm]{f1b.ps}
\caption{(Left) Data for XIS0, 1 and 3 are shown in black, red, and green, respectively. (Right) Zoom-in on Fe lines region, shown on a linear scale. 
 \label{fig:xis}}
  \end{center}
\end{figure*}

\begin{figure*}
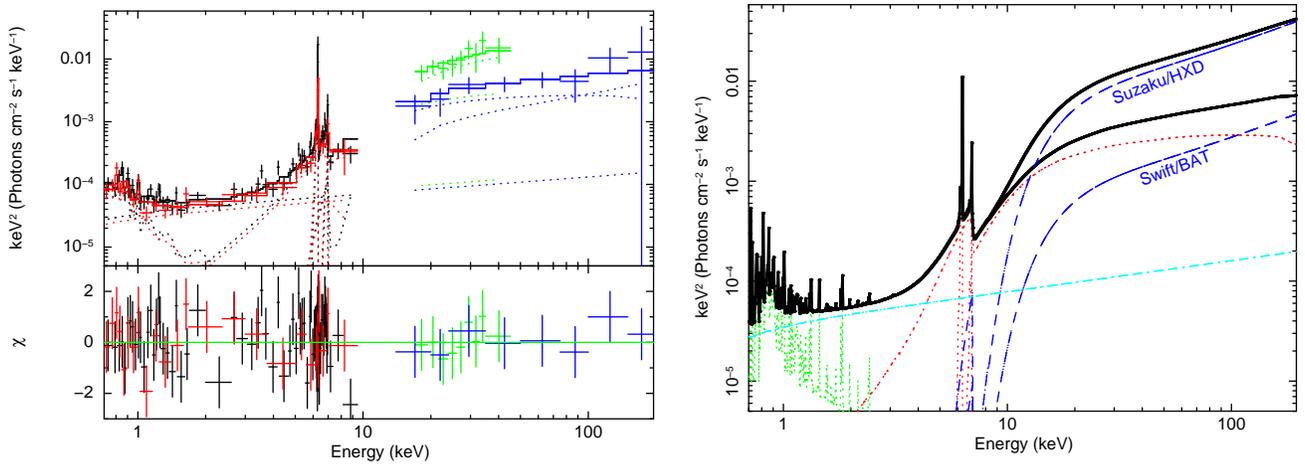

  \begin{center}
    \includegraphics[angle=270,width=8.6cm]{f2a.ps}
    \includegraphics[angle=270,width=8.5cm]{f2b.ps}
\caption{(Left) Best-fitting unfolded model M and data for front-illuminated (XIS0+3) CCDs in black and back-illuminated XIS1 are in red, while those for HXD/PIN and BAT are shown in green and blue, respectively. 
(Right) Corresponding model components: IPL (blue dashed), {\sc mytoruss + mytorusl} (red dotted), SPL (cyan dash-dot) and {\sc apec} (green dotted).
 \label{fig:spec}}
  \end{center}
\end{figure*}

\section{X-ray fitting}

For modeling the broadband data, the soft emission below 2 keV was parametrized with a hot thermal plasma {\sc apec} component \citep{apec}, although alternatives are discussed in Section~\ref{sec:discussion_soft}. The underlying continuum component is an absorbed intrinsic power-law (hereafter, IPL) with photon-index $\Gamma_{\rm IPL}$. Furthermore, a fraction ($f_{\rm SPL}$) of the IPL scattered into the line-of-sight (hereafter, SPL) from an optically-thin medium, was found to be necessary. Emission lines were modeled as Gaussian profiles at a fixed systemic redshift $z$=0.016285\footnote{http://ned.ipac.caltech.edu}. Fixed Galactic \nh=4.1$\times$10$^{20}$ cm$^{-2}$ was also included, based upon \hi\ measurements along the line-of-sight \citep{dickeylongman90}. 
The two front-illuminated (fi) CCDs XIS0 and 3 were combined for fitting, and the energy ranges of 1.7--1.9 and 2.1--2.3 keV were ignored due to instrumental calibration uncertainties related to Si and Au edges. All spectral fitting was carried out using {\sc xspec} v12.8 \citep{xspec} and uncertainties are quoted at the 90\% level. 

\subsection{Absorbed power-law plus \pexrav\ : model P}

We next added in a \pexrav\ \citep{pexrav} component characterizing \lq pure reflection\rq\ (i.e. negative-valued model parameter $R$ in \xspec\ terminology). The reflected component was taken to be unabsorbed, while photoelectric absorption ({\tt phabs}) and Compton down-scattering ({\tt cabs}; including Klein-Nishina corrections at high energies) due to a single heavy obscurer were introduced for the IPL. This model is referred to as Model P. HXD data cross-normalization was fixed at 1.16 relative to fi\footnote{http://heasarc.gsfc.nasa.gov/docs/suzaku/analysis/abc/node8.html\# SECTION00872000000000000000}, while that for the back-illuminated CCD was left free.

\subsection{\mytorus\ : model M}

The \pexrav\ model assumes reflection off a slab with infinite optical depth. More sophisticated treatment of reflection, including a classical circumnuclear torus geometry and self-consistent inclusion of fluorescence emission was carried out with the \mytorus\ model \citep{mytorus}. All \lq minimal\rq\ table model components were included. These comprise: i) the distortion of the zeroth-order continuum due to obscuration by the torus, ii) Compton-scattering off gas in the torus ({\sc mytoruss}), and iii) associated fluorescence line emission ({\sc mytorusl}). The normalizations of these three components were tied to that of IPL. 
\mytorus\ assumes no high-energy cutoff up to a termination energy of 500 keV, and has precomputed table model grids for \nh\ up to 10$^{25}$ cm$^{-2}$. Fe K$\beta$ (7.06 keV) is included by default in the model, reducing the significance of Fe {\sc xxvi}. \mytorus\ allows decoupling the los obscuring column (\nh(los)) from that responsible for the scattered continuum and fluorescence (\nh(scatt)). This model is referred to as model M.

\subsection{Accounting for long-term variability}

As we will discuss in detail later, the \suzaku/HXD data show a higher hard X-ray flux level than \swift/BAT. Under the assumption that variability is dominated by the power-law, we introduced a further cross-normalization constant ({\sc const}$_{\rm IPL}$) in both models P and M to account for different intrinsic (i.e. IPL) source fluxes measured by \suzaku\ and \swift. In this scenario the absolute flux of the reflection and scattered components do not vary, with their normalizations being tied to the same value as the IPL between all instruments. The XIS regime, being reflection-dominated, is not greatly sensitive to changes in the IPL, though it is treated self-consistently throughout.

\section{Results}

The \suzaku\ XIS spectra are shown in Fig.~\ref{fig:xis}. A flat continuum above 2 keV and prominent emission due to Fe fluorescence are seen. Characterizing the XIS spectrum with an absorbed IPL alone results in an extremely-flat $\Gamma_{\rm IPL}^{\rm XIS}$=0.6$_{-0.4}^{+0.5}$, but with an unacceptable goodness-of-fit statistic. These properties clearly imply the presence of heavy obscuring and reflecting matter in the source. 

With regard to the broadband fits using models P and M, acceptable fits were found in both cases with the parameters reported in Table~\ref{tab:fits}. The determined line-of-sight (los) \nh\ and its confidence limits lie well above 10$^{24}$ cm$^{-2}$ implying CT obscuration. In model M, \nh(los) is larger than \nh(scatt). The underlying power-law appears to show clear variability between \suzaku\ and \swift, characterized by {\sc const}$_{\rm IPL}$. The IPL high energy cutoff ($E_{\rm cutoff}$) remains unconstrained and was fixed at 500 keV in model P. The inclination angle of the reflector ($\theta$) was left free and is also unconstrained. 

The equivalent width (EW) of Fe K$\alpha$ on the observed continuum is $\sim$1.1 keV. At least one additional emission line is required, with centroid energy close to that of redshifted Fe {\sc xxvi} (6.97 keV) and EW(Fe {\sc xxvi})$\approx$0.3 keV. Adding a third Fe line (e.g. due to Fe K$\beta$) does not provide a significant improvement. 

The resultant fits of the two models (Table~\ref{tab:fits}) show that model M is marginally better than model P, though both are acceptable statistically. The unfolded Model M is shown in Fig.~\ref{fig:spec}.

\begin{table*}
{\scriptsize
  \begin{center}
  \caption{\xspec\ fits to the X-ray spectrum of \eso565.\label{tab:fits}}
  \begin{tabular}{m{5cm}m{2cm}m{2cm}m{3.5cm}}
    \hline 
                                 &    AGN Model P$^*$ & AGN Model M$^*$ &        \\
    Component and Parameters     & ({\sc pexrav})    & ({\sc mytorus})     & ~~~~~Units\\
    \hline                                                  
  ${\rm APEC}\left\{
  \begin{tabular}{m{4cm}m{2cm}m{2cm}m{2cm}}
    $kT$                &    0.69$_{-0.10}^{+0.07}$      &    0.67\p0.09        & keV\\
    Abundance           &    0.14$_{-0.18}^{+u}$      &    0.3$^\ddag$        &\\
    Norm                &    15.3$_{-9.1}^{+18.9}$         &   7.6\p1.6         &$\times$10$^{{-5}\ \diamondsuit}$\\
 $F_{0.5-2}^{\rm obs}$  &    1.42$_{-0.36}^{+0.03}$      &  1.44$_{-0.46}^{+0.03}$        & $\times$10$^{-13}$~erg~s$^{-1}$~cm$^{-2}$\\
 $L_{0.5-2}^{\rm APEC}$($\dag$) &    6.3                  &     5.0       & $\times$10$^{40}$ erg s$^{-1}$\\
                        &                              &            &\\
  \end{tabular}
 \right.$\\
  ${\rm AGN\ Continuum}\left\{
  \begin{tabular}{m{2.7cm}m{2cm}m{2cm}m{2cm}}
     \nh({\rm los})                &      2.1$_{-0.5}^{+1.5}$        &  4.4$_{-1.5}^{+u}$          & $\times$10$^{24}$ cm$^{-2}$\\
     \nh({\rm scatt})              &      --        &  1.0$_{-0.4}^{+1.3}$          & $\times$10$^{24}$ cm$^{-2}$\\
  $\Gamma_{\rm IPL}$    &      1.69$_{-0.29}^{+0.31}$      &   1.72$_{-0.27}^{+0.41}$        & \\
$\theta$             &      81$_{-u}^{+u}$ &      67$_{-42}^{+14}$       &  deg\\
$E_{\rm cutoff}$       &   500$^\ddag$          &    --       & keV \\
 Norm$_{\rm IPL}$       &      $2.8_{-2.6}^{+28}$      &     4.2$_{-2.2}^{+14}$      & $\times$10$^{-3}$~ph~s$^{-1}$~cm$^{-2}$~keV$^{-1}$\\
     $\Gamma_{\rm SPL}$ &      =$\Gamma_{\rm IPL}$      &   =$\Gamma_{\rm IPL}$        & \\
                           &             &            &\\
 $F_{2-10}^{\rm obs}$   &      6.5$_{-4}^{+10}$          &   7.1$\pm$0.8         & $\times$10$^{-13}$~erg~s$^{-1}$~cm$^{-2}$\\
 $L_{\rm XIS\ (2-10)}^{\rm IPL}$($\dag$)   &      3.3         &   6.2         & $\times$10$^{43}$ erg s$^{-1}$\\
 $L_{\rm BAT\ (14-195)}^{\rm IPL}$($\dag$)  &  2.5        &   3.4         & $\times$10$^{43}$ erg s$^{-1}$\\
 $L_{\rm HXD\ (10-100)}^{\rm IPL}$($\dag$)  &  8.6        &   15.3         & $\times$10$^{43}$ erg s$^{-1}$\\
                        &             &            &\\
     {\sc const}$_{\rm IPL}^c$ & 4.6$_{-2.2}^{+52}$     &    6.0$_{-2.7}^{+127}$       &\\
 $L_{\rm 2-10}^{\rm IPL}$($\dag$)   &      0.7         &   1.0         & $\times$10$^{43}$ erg s$^{-1}$\\
       $R$              &      --0.9$_{-u}^{+1.9}$ &     --      & \\
   $f_{\rm SPL}$        &      1.0$_{-0.9}^{+35}$     &    0.9$_{-0.7}^{+0.6}$ & $\times$10$^{-2}$    \\
  \end{tabular}
 \right.$\\
  ${\rm Lines}\left\{
  \begin{tabular}{m{4.25cm}m{2cm}m{2cm}m{2cm}}
$E_{\rm obs}$   &   6.28\p0.01         &   --        & keV \\
$\sigma$   &   0.001$^\ddag$      &    $^m$        & keV\\
  Norm     &   1.09\p0.12        &            & $\times$10$^{-5}$~ph~s$^{-1}$~cm$^{-2}$ \\
EW           &     1.1\p0.6       &    1.2        & keV\\
$E_{\rm rest}$  &   6.38\p0.01         &   $^m$        & keV \\
$\Rightarrow$ ID=Fe K$\alpha$ &&&\\
                             &             &            &\\
$E_{\rm obs}$ &   6.84$\pm$0.07      &   --       & keV \\
      $\sigma$ & 0.001$^\ddag$       &      --      & keV \\
      Norm  & 2.1\p1.2      &   --         & $\times$10$^{-6}$~ph~s$^{-1}$~cm$^{-2}$ \\
EW    &   0.20$_{-0.17}^{+0.20}$      &    --        & keV\\
$E_{\rm rest}$ &  6.95$\pm$0.07      &   --       & keV \\
$\Rightarrow$ ID=Fe {\sc xxvi} &&&\\
                            &             &            &\\
EW$_{\rm K\beta}$ &  --       &    0.2        & keV\\
                                &             &            &\\
  \end{tabular}
 \right.$\\
                           &             &            &\\
  \begin{tabular}{m{5.2cm}m{2cm}m{2cm}m{2cm}}
     $\chi^2$/dof  & 96/103     &   95/108         &\\
   \end{tabular}
                   &             &            &\\
\hline
  \end{tabular}\\
~\\
$^*$Model P: {\sc const * phabs[}$\mapsto$$N_{\textrm {\tiny H}}^{\textrm {\tiny Gal}}${\sc ] ( apec + pexrav + phabs*cabs*const$_{\textsc{\tiny IPL}}$*pow[}$\mapsto${\sc ipl] + const[}$\mapsto$$f_{\rm SPL}${\sc ]*pow[}$\mapsto${\sc ipl] + gauss($\times$2) ).}\\
$^*$Model M: {\sc const * phabs[}$\mapsto$$N_{\textrm {\tiny H}}^{\textrm {\tiny Gal}}${\sc ] ( apec + ({\sc *etable}\{{\tt mytorus\_Ezero\_v00.fits}\}*const$_{\textsc{\tiny IPL}}$*pow[}$\mapsto${\sc ipl]} + {\sc atable}\{{\tt mytorus\_scatteredH500\_v00.fits}\} + {\sc atable}\{{\tt mytl\_V000010nEp000H500\_v00.fits}\} + {\sc const[}$\mapsto$$f_{\rm SPL}${\sc ]*pow[}$\mapsto${\sc ipl]} ). \\
$^*$In both models, the first {\sc const} refers to the instrumental cross-normalization. Galactic absorption $N_{\textrm {\tiny H}}^{\textrm {\tiny Gal}}$ is fixed to 4.06$\times$10$^{20}$ cm$^{-2}$. The normalizations for the table components are tied to IPL.\\ 
$^u$unconstrained.\\
$^\ddag$fixed.\\
$^\diamondsuit$\apec\ norm quoted in units of 10$^{14}$ cm$^{-5}$.\\
$^c$HXD IPL normalization relative to BAT.\\
$^m$Fluorescence in {\sc mytorus} model includes Fe K$\alpha$ and K$\beta$ assuming no velocity shift. A smoothing width of $\sigma$=1 eV was assumed.\\
$\dag$Absorption-corrected luminosities for the {\sc apec} and AGN components are quoted after setting all other model components to zero. $L_{\rm 2-10}$ is to be regarded as the long-term 2--10 keV IPL power and is equal to $L_{\rm XIS\ (2-10)}$/{\sc const}$_{\textsc{\tiny IPL}}$.
  \end{center}
}
\end{table*}

\section{Discussion}

Using \suzaku, we have carried out the first pointed X-ray study of \eso565. In this section, we discuss some salient points of our analysis. 

\subsection{X-ray spectrum below 10 keV}
\label{sec:discussion_soft}

The soft band below about 2 keV has been modeled with contributions from a hot interstellar thermal plasma and the SPL which has a scattering fraction ($f_{\rm SPL}$) of a few per cent, as has been inferred from modeling low spectral resolution (CCD) data of Seyfert galaxies \citep[e.g. ][]{turner97_data, cappi06}. Early-type galaxy halos and sources dominated by starburst activity typically show thermal emission from diffuse plasma with temperatures of $\ltsim$1 keV (e.g. \citealt{forman85, done03, konami12}). On the other hand, high resolution grating spectra of bright nearby Seyferts, including other CT-AGN, have instead shown that the soft X-ray regime may be dominated by emission from gas photoionized by the nucleus \citep[e.g. ][]{sako00, kinkhabwala02, iwasawa03, armentrout07, guainazzi07}. Distinguishing between these various scenarios will require higher spectral resolution data than we possess at present, which should be possible using the micro-calorimeter on board \astroh\ \citep{astroh12}. 

With regard to the SPL, the luminosity in this component is $\sim$10$^{41}$ erg s$^{-1}$. This is low enough that some part of the scattered emission may arise as the integrated emission from point sources other than the AGN within \eso565\ (e.g. X-ray binaries). \suzaku\ is unable to disentangle this component spatially, but this should be possible with the \c\ satellite. 

Over the 4--10 keV regime, the observed emission is found to be reflection-dominated (RD). With its high \nh(los) and strong Fe line, \eso565\ joins the ranks of well-known local CT-AGN with strong Fe lines such as NGC 1068 \citep[see, e.g., the compilation by ][]{levenson02}. 

The variability observed between HXD and BAT above 10 keV (discussed further in \S~\ref{sec:hxdbat}) also has consequences for the lower energy range. It is important to remember that we have assumed the variability to be dominated by the IPL normalization alone, and that the absolute fluxes of the reflection and scattering components do not change because they ought to respond on long timespans. This means that any fit variables defined relative to the IPL (i.e. $f_{\rm SPL}$, $R$) also change depending upon the dataset considered. Table~\ref{tab:fits} states values relative to the long-term IPL flux probed by \swift/BAT. So when considering the values of these variables in the \suzaku\ data alone, these will effectively be lower by a factor of {\sc const}$_{\rm IPL}$. Similarly, in Table~\ref{tab:fits}, we have distinguished between the absorption-corrected IPL luminosities $L_{\rm XIS\ (2-10)}$ and $L_{\rm 2-10}$. $L_{\rm XIS\ (2-10)}$ is the absorption-corrected 2--10 keV luminosity for the XIS consistent with the HXD flux level (the two instruments observed the source simultaneously on the same date). On the other hand, $L_{\rm 2-10}$ is lower by a factor of {\sc const}$_{\rm IPL}$, and may be considered the long-term average 2--10 keV IPL luminosity based upon the \suzaku/XIS and \swift/BAT datasets. These variables are collected and tabulated together with {\sc const}$_{\rm IPL}$ in Table~\ref{tab:fits}.

The inferred long-term absorption-corrected IPL luminosity for models P and M ranges over $L_{2-10}$$\sim$(0.7--1)$\times$10$^{43}$ erg s$^{-1}$. Mean X-ray bolometric corrections and Eddington fractions of the \swift/BAT AGN sample have been studied by \citet{vasudevan10}, who showed that this sample has a comparatively low Eddington ratio distribution overall \citep{vasudevan10}. For a typical bolometric correction ($L_{\rm Bol}$:$L_{2-10}$)$\sim$10--30, the mean $L_{\rm Bol}$ of ESO~565--G019 lies within the range of $\sim$(0.7--3)$\times$10$^{44}$ erg s$^{-1}$. If the accretion luminosity of the source is 10~\%\ of the Eddington value on average, this would imply a supermassive black hole mass range of (0.5--2)$\times$10$^7$~\Msun. 

\subsection{Fe complex}

The Fe K$\alpha$ line is strong, with EW(Fe K$\alpha$) in excess of 1 keV with respect to the observed continuum. With respect to the reflection continuum alone, we find EW(Fe K$\alpha$)$\approx$1.8 keV (model P) and 1.4 keV (model M). Fe line fluorescence can, in principle, constrain the geometry of the reflecting matter. The observed range of EWs is certainly consistent with the line and reflection continuum originating in the same location, presumably the torus \citep[][]{matt96_felines}. With regard to the inclination angle of the reflector, Table~\ref{tab:fits} suggests highly inclined $\theta$ values. In the parameter space of torus reflection models \citep[e.g. ][]{matt96_felines, levenson02}, higher inclinations favor larger Fe line equivalent widths, qualitatively consistent with our findings. But we stress that uncertainties on $\theta$ are not negligible, and this parameter remains unconstrained in model P.

Down-scattering of emission line photons also results in a characteristic hump known as the Compton shoulder (CS) redward of the rest-frame line centroid. The strength of this feature can be a sensitive diagnostic of the column density (\citealt{matt02}). CS signatures have been found in several CT AGN studied by \suzaku\ (e.g. \citealt{itoh08, awaki08}). A single down-scattering results in a maximum energy decrement of $\approx$160 eV, which is close to the spectral resolution of the XIS CCDs. In fact, Table~\ref{tab:fits} shows that the rest frame energy of the Fe K$\alpha$ line in ESO~565--G019 measured using a Gaussian profile in model P is slightly lower than the expected value of 6.4 keV, which is likely to be caused by a CS blended with the line.

A CS is included self-consistently within model M through \mytorus. In order to independently check for its existence, we added a CS component to model P, parametrizing it using {\tt xsgacp} \citep{madejski06,illarionov79}. We fixed the electron temperature of the scattering medium to a nominal value of 1~eV, and the initial line energy to rest frame 6.40~keV. This resulted in a decrease of $\chi^2\sim 2$ for the same number of degrees of freedom as in Table~\ref{tab:fits} (as the neutral K$\alpha$ line energy is now fixed). The CS component contains a fraction $0.28_{-0.22}^{+0.28}$ of the narrow line photons. The detection of the CS is significant and consistent with the expectation of $\sim 0.2$ for reflection from Compton thick material \citep{matt02}, though uncertainties are large.

\subsection{Hard X-ray spectrum and long-term variability}
\label{sec:hxdbat}

Fitting over the full energy range available with \suzaku/XIS and \swift/BAT suggests that the IPL dominates above $\sim$100 keV in the BAT band, and $E_{\rm cutoff}$ should lie beyond the BAT energy sensitivity range. The fitted $\Gamma_{\rm IPL}$ values are consistent with the median of the distribution of hard X-ray photon-indices for radio-quiet Seyferts \citep{dadina08}. 

\nh(los) is found to be larger than \nh(scatt) in model M, which may be a result of clumpiness in the absorber, as supported by a wide range of observational and theoretical work on AGN tori in general (e.g. \citealt{hoenig13_torusmodels} and references therein). But with the IPL dominating only in the two highest energy BAT bins, these data do not probe the underlying continuum in detail. This is an important limitation of reflection-dominated spectra, and leads to various degeneracies (e.g. \nh(los) being unconstrained at the high end; large uncertainties on {\sc const$_{\rm IPL}$}), and an even more reflection-dominated hard-band spectrum cannot be ruled out. 

The observed \suzaku/HXD flux ($F_{\rm 15-40\ keV}$$\approx$1.4$\times$10$^{-11}$ erg s$^{-1}$ cm$^{-2}$) lies above that seen in \swift/BAT by a factor of $\approx$3--4 over the same energy range. Systematic uncertainties are important when accounting for the HXD/PIN background reproducibility. Typical earth-occultation data systematic errors are $\sim$3.8\%\footnote{http://heasarc.gsfc.nasa.gov/docs/astroe/prop\_tools/suzaku\_td/node12.html\# SECTION001251100000000000000}. Including an increased PIN background at this level reduces the HXD excess with respect to BAT to a factor $\sim$2. 

These differences cannot be caused by confusion and contamination of the HXD data, because the nearest neighbor in the BAT 70-month catalog is PKS~0921--213 at a distance of 2$^\circ$.6, well outside the collimated $\sim$34\arcmin--sized PIN fov. 

The BAT data represent a long-term average of the source flux between Dec 2004 and Sep 2010. \suzaku\ observations were carried out 20 months later, so hard X-ray variability may explain the difference. Reflection and scattering are generally thought to occur on $\sim$pc-scale toroidal clouds. Rapid variations are naturally associated with the IPL (as we have modeled). The lower limit of the IPL variation in this scenario is at least a factor of $\sim$2. 

It should be noted that we cannot completely rule out strong variations of the hard band Compton hump originating in an inner disc reflection component (e.g. due to light-bending). This possibility can be tested in follow-up X-ray monitoring to search for rapid changes in the reflection continuum and Fe line. 

\subsection{Multi-wavelength comparisons}
\label{sec:multiwavelength}

\eso565\ has not been followed up in detail at any wavelength previously, but it has been covered in all-sky surveys in the mid-infrared (MIR), a regime that can provide efficient selection of highly obscured AGN. 
For example, the study of \citet{severgnini12} finds a MIR number density of CT-AGN that is 2--4 times higher than present hard X-ray surveys, though some correction for IR star-formation contamination is necessary. 

The \akari\ satellite carried out an all-sky survey in 2006--2007 in multiple mid- and far-infrared bands \citep[][ and references therein]{akari_allsky}. The reported fluxes of \eso565\ at 9 and 90~\micron\ are 77\p13 mJy and 2.90\p0.11 Jy, respectively. Combining these with the observed BAT luminosity, we find that \eso565\ lies close to other Compton-thick sources NGC~6240 and NGC~5728 in the MIR:hard-X-ray \lq color-color\rq\ plot of \citet[][ their Fig. 5]{matsuta12}, but not significantly different from the overall distribution of Seyfert colors. This relatively-high 90:9~\micron\ flux ratio suggests the presence of large quantities of cool dust which is well-known for NGC~6240 and NGC~5728 \citep[e.g. ][]{rieke85, schommer88}, but atypical for an early-type galaxy like \eso565 (this is discussed further in section~\ref{sec:morph}).

The observed (absorbed) 2--10 keV X-ray luminosity in the \suzaku\ data is $L_{2-10}$$\sim$4$\times$10$^{41}$ erg~s$^{-1}$, and the modeled absorption-corrected power is 1--2 orders of magnitude higher depending upon which dataset is considered. X-ray modeling of CT-AGN can be highly sensitive to the geometry of the absorber and reflector when the underlying continuum cannot be probed clearly \citep[e.g. ][]{mytorus}. It is thus useful to compare against other diagnostics of the intrinsic power, as an independent check of the appropriateness of the X-ray modeling. 

Few emission line spectroscopic measurements of the source have been published. Considering photometric data, it has been shown that there is a close correlation between the monochromatic 12\micron\ continuum torus luminosity and intrinsic 2--10 keV power for local Seyferts \citep[e.g. ][]{horst08, asmus11}. The best constraints come from isolating the torus spatially in high angular resolution imaging, but such data are not available for \eso565. The source is detected in the \wise\ all-sky survey\footnote{http://wise2.ipac.caltech.edu/docs/release/allsky/} \citep{wise}. Its W3 band (12 \micron) flux is $F$(W3)=96\p11 mJy, or $\lambda L_{\lambda}$(W3)=1.6$\times$10$^{43}$ erg s$^{-1}$. Using Eq. 2 from \citet{g09_mirxray} then implies $L_{2-10}^{\rm predicted}$=1.1(\p0.2)$\times$10$^{43}$ erg s$^{-1}$. This is smaller than the instantaneous IPL power $L_{\rm XIS\ (2-10)}$ measured in the XIS, but closely matches the inferred long-term intrinsic power estimates $L_{2-10}$ (Table~\ref{tab:fits}). The close match is encouraging because it is the long-term power which is relevant from the point-of-view of a pc-scale dust torus. Still, we caution that this is an approximate cross-check only, because various caveats are unaccounted for (e.g., \wise\ cannot resolve the nucleus with its nominal 6\farcs 5 beam; and, the W3 bandpass is broad enough to be affected by various emission and absorption features). In addition, low-level anisotropy in the MIR emission is expected in clumpy torus models \citep[e.g. ][]{hoenig11_anisotropy}. Broadband data that have better direct sensitivity to the IPL will allow more detailed multi-wavelength comparisons.

\subsection{Implications for future surveys} 

Present estimates of the contribution of CT-AGN to the cosmic X-ray background (CXB) flux are of order 10~\%, with the largest uncertainty being the prevalence of RD sources \citep[e.g. ][]{treister09}. A wide range of RD-AGN contributions are allowed, in principle \citep{g07}. At the flux levels probed by current hard X-ray missions, only a tiny fraction of the hard CXB is resolved into point sources, while CT-AGN are inferred to constitute $\sim$10--20\% of the overall AGN number density \citep[e.g. ][]{sazonov07a, burlon11, ricci11, vasudevan13_58m}. 

\eso565\ was detected in the BAT 39-month catalog, but not earlier ones \citep{cusumano10_39, tueller10}. Thus, our work suggests that other RD-AGN are waiting to be uncovered by hard X-ray selection at fainter fluxes. The new generation of hard X-ray focusing missions \nustar\ and \astroh\ will obtain detailed hard-band spectra of sources like \eso565\ and test the cosmological significance of this population \citep{ballantyne11, astroh12}. 

\subsection{CT-AGN in an early-type galaxy}
\label{sec:morph}

\begin{figure*}
  \begin{center}
    \includegraphics[angle=0,height=8.5cm]{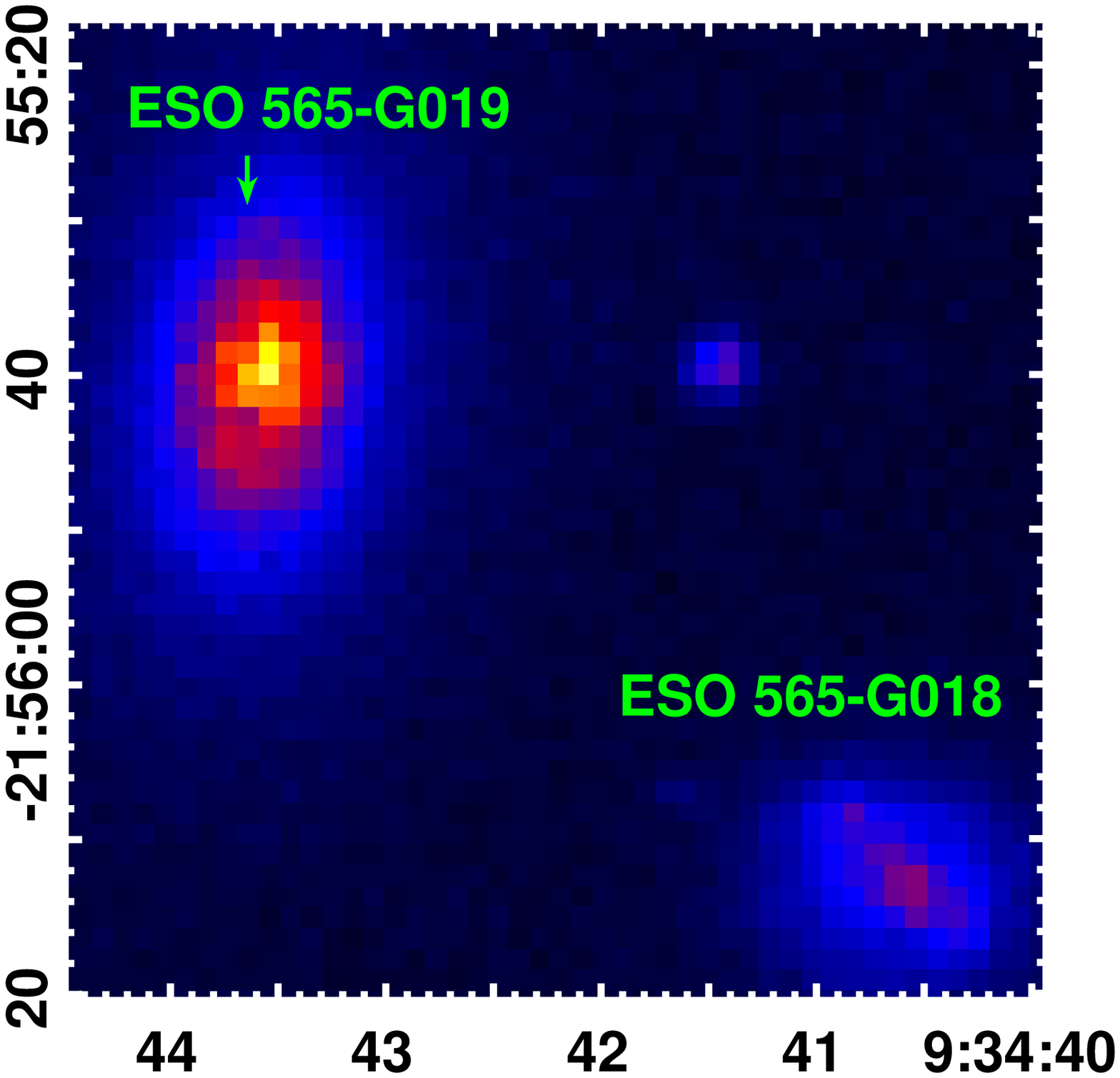}
    \includegraphics[angle=0,height=8.5cm]{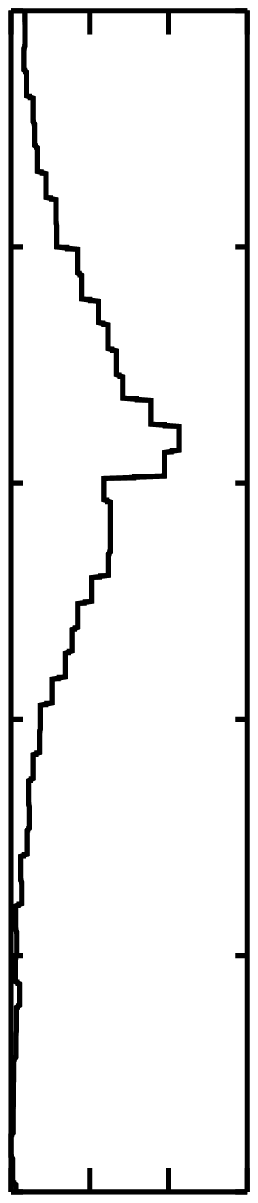}
\caption{Archival $R$-band image from the ESO-LV catalog of galaxies {\tt (archive.eso.org/wdb/astrocat/esolv.html)}. The core of ESO~565--G019 appears asymmetric, with a slight excess skewed to the north and an apparent flux deficit to its south--east. This is best seen in the vertical profile cut on the right, drawn through the image pixel column containing the arrow. Linear color scaling has been used.
\label{fig:esor}}
\end{center}
\end{figure*}

The host galaxy morphology of \eso565\ was first classified as type E in the ESO/Uppsala survey by \citet{lauberts82}. A digitized image from this survey is shown in Fig.~\ref{fig:esor}. The Third Reference Catalog of Bright Galaxies (RC3) also reported the same classification, but with an extra \lq ?\rq\ flag indicating doubt \citep{rc3}. The most recent report, based upon a by-eye classification of Digitized Sky Survey plates, is of type S0 \citep{huchra12}. Diffraction-limited optical imaging with modern telescopes has not been carried out for this object. Seeing-limited CCD imaging has been analyzed by \citet{prugniel98} and \citet{alonso03}. Through profile-fitting, \citeauthor{prugniel98} find a S\'{e}rsic index $n$=3.0, suggesting that this source is unlikely to host a pseudo-bulge which typically have $n$$\ltsim$2 (see discussion and references in Section 3.1 of \citealt{vaghmare13}). \citeauthor{alonso03} obtain an excellent fit by decomposing the total galaxy light profile into a bulge-dominated system plus a disk, with a bulge/disk ratio of 2.56, a typical value for their sample of early-type galaxies. In summary, the host is known to be an early-type bulge-dominated system, but the exact classification remains uncertain. 

Only about 20 bona fide CT AGN have been identified and studied in detail in the local universe. They are typically radio-quiet systems hosted in spiral galaxies, with only a handful hosted in galaxies with Hubble types S0\footnote{ESO~138--G001 \citep{piconcelli11}, Mrk 3 \citep[e.g. ][]{awaki08}, and ESO~323--G032 \citep{comastri10}}, and none in type E (see, e.g., Table 1 of \citealt[][]{goulding12} and the compilation of \citealt{dellaceca08}). LINERs are invariably X-ray--faint, precluding secure CT classification. In the distant universe, morphological classification is more difficult. Based upon the HST imaging of distant AGN hosts presented in \citet{gabor09}, \citet{trump11} has shown there is a strong preference for obscured AGN to lie in disky and irregular galaxies (though their sample does not include CT AGN). Heavily obscured {\em radio} galaxies can be associated with spheroidal systems, but these are often hosted in clusters where interactions may drive gas to the nucleus. \eso565\ is neither radio-loud, nor does it lie in a cluster. There are similarities in some respects between \eso565\ and the nearby FR I prototype Cen A \citep{israel98}, in that both are early-type systems with dust. One important difference is that the obscuring column density of gas to the nucleus of Cen A is significantly lower and Compton-thin \citep[\nh$\sim$10$^{23}$ cm$^{-2}$;][]{risaliti02_bright}.

\begin{figure*}
  \begin{center}
    \includegraphics[angle=0,height=8.5cm]{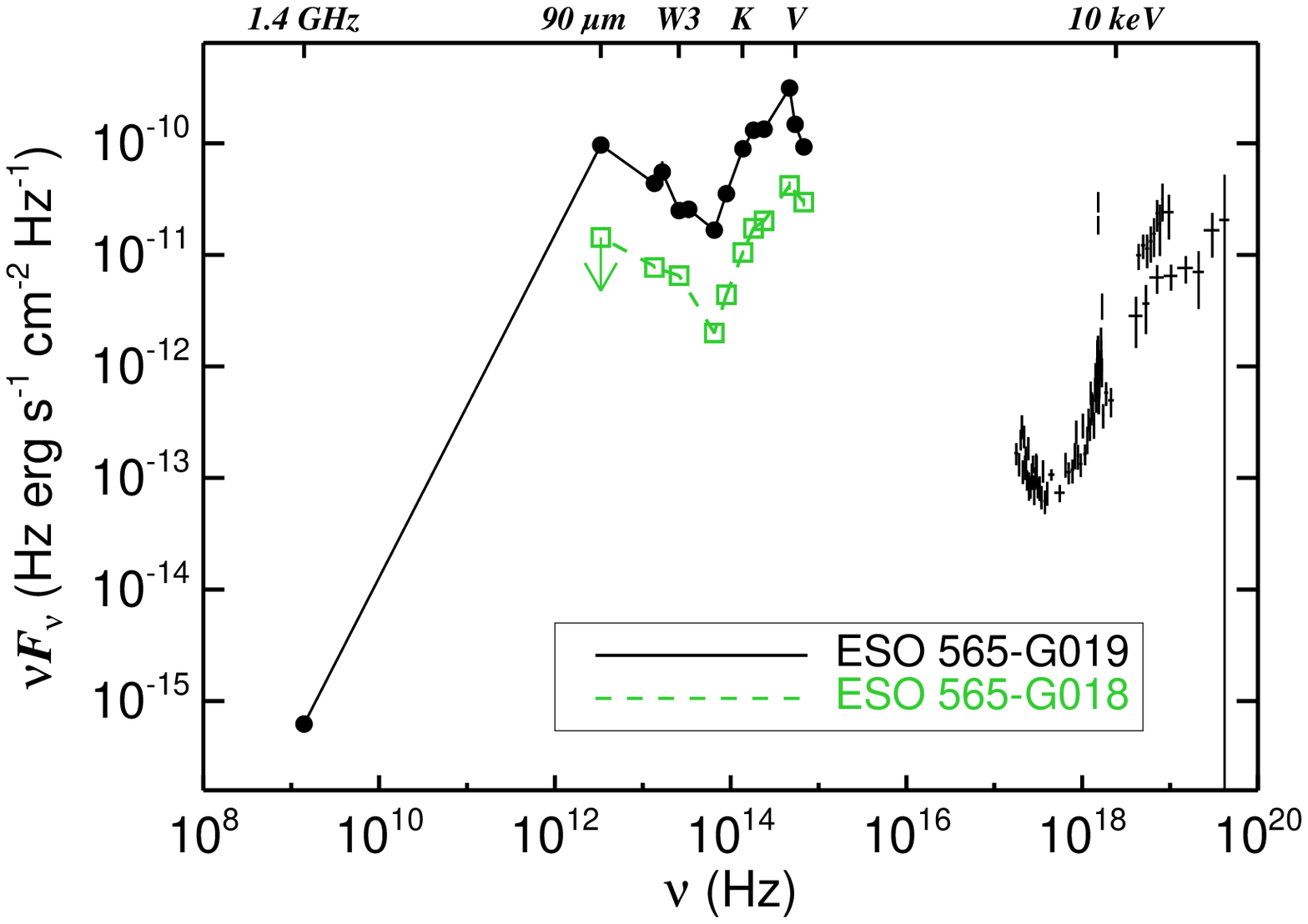}
\caption{Broadband SED of ESO~565--G019 in black including our \suzaku+\swift\ X-ray unfolded model M, together with the following archival measurements: 1.4 GHz radio data from the NRAO VLA sky survey \citep{condon98}; \akari\ 90~\micron\ far-infrared data from \citet{akari_allsky_fis}; \akari\ 18 and 9~\micron\ data from \citet{akari_allsky}; 22, 12, 4.6 and 3.4~\micron\ MIR data from the \wise\ all-sky data release \citep{wise}; $Ks$, $H$ and $J$ band near-IR data from 2MASS, as measured in 14$\arcsec$$\times$14$\arcsec$ aperture \citep{2mass}; $R$ and $B$ Cousins filter photometry from the ESO-LV survey showing $R_{25}$ and $B_{25}$ standard aperture measurements \citep{lauberts82}; and $V$-band ($V_T$) data from RC3 \citep{rc3}. In green, data for ESO~565--G018 are shown. These are drawn from the same sources, where available. This source is undetected with \akari, and the nominal 80\%\ completeness limit of 0.43~Jy is plotted as an upper-limit at 90~\micron\ \citep{akari_allsky_fis}.
\label{fig:sed}}
\end{center}
\end{figure*}

The question of the origin and abundance of dust in early-type galaxies remains to be settled. Detailed photometric analyses have revealed the presence of extended dust lanes, filaments or disks in {\em most} systems with high quality observational data \citep[e.g. ][]{vandokkumfranx95, tran01_ellipticals}. Recent \herschel\ observations also reveal that dust is prevalent in early-type galaxies \citep{smith12_dustetgs}. ESO~565--G019 is a warm infrared galaxy with a total IR luminosity $L_{\rm IR}=10^{10.47}$ \Lsun\ and was prominently detected by \iras\ in all its observing bands \citep{kewley01, strauss92} showing that cool dust is present in the host on large scales. This can be seen in the broadband SED shown in Fig.~\ref{fig:sed}. Using the total IR luminosity and its known relation to star formation rate (SFR) relation \citep[][Eq. 4]{murphy11}, we find SFR(\eso565)$\sim$3--4 \Msun\ yr$^{-1}$ depending upon whether one uses the \akari\ or \iras\ data. This SFR is a few times higher than that of the Milky Way, and about an order of magnitude larger than in typical early-type systems \citep{crocker11}. In order to estimate the mass of this extended cool dust component, we use Eq.~1 of \citet{smith12_dustetgs} and their assumptions about the dust properties, in conjuction with the observed 100~\micron\ \iras\ flux of ESO~565--G019. The main uncertainty is the dust temperature ($T_{\rm d}$), because the reported far-infared fluxs do not probe beyond 90--100~\micron. A $T_{\rm d}$=40 K blackbody can describe the 60--100 \micron\ regime well, though we caution that even cooler dust may well be present. This yields a total cool dust mass of $\sim$5$\times$10$^6$~\Msun\ for ESO~565--G019, a value which lies at the upper end of the range of dust masses found by \citet{smith12_dustetgs} and about an order of magnitude larger than the mean dust mass for their full sample. Assuming $T_{\rm d}$=30 K would increase the mass by a further factor of 3.

But high columns of hot {\em nuclear} dust and associated obscuring gas in typical early-type systems remain rare. Only about 3\%\ of BAT AGN are hosted in ellipticals \citep{koss11}, with none identified as being CT thus far. Thus, confirmation of the exact morphological class of \eso565\ is important. If it turns out to be of E-type, it would be the first of its kind.

\subsection{Possible triggers of nuclear activity}
\label{sec:trigger}

\subsubsection{Interaction with ESO~565--G018}

\eso565\ is a member of a wide pair with ESO~565--G018 \citep{lauberts82, reduzzi95}, separated by a projected distance of $\approx$17~kpc. Fig.~\ref{fig:esor} shows that ESO~565--G018 appears to be elongated along their separation line. If the nuclear activity in \eso565\ is a result of neighbor interaction, this is not accompanied by significant morphological disruption of its host. 

Simulations of galaxy mergers show that pronounced star formation and nuclear activity can follow close encounters between the merging pair (e.g. \citealt{dimatteo05}). This means that ESO~565--G018 and ESO565--G019 have likely undergone at least one close passage already. We have discussed the elevated SFR in \eso565\ as inferred from its high total IR luminosity. Its neighbor ESO~565--G018 is about a factor of 10 fainter in the optical and IR, and has remained undetected with both \iras\ and \akari\ in the far-IR. Fig.~\ref{fig:sed} compares the broadband SEDs of the two galaxies. So, if a close interaction is ongoing, it appears to have triggered elevated star formation in one system but not the other. 

In this respect, the interaction may be similar to that of M81 and M82 -- the nearest galaxy pair where a close interaction has led to a well-studied nuclear starburst and pronounced morphological distortion in M82 (e.g. \citealt{g11_m82} and references therein) but not its neighbor. A perigalactic approach of $\sim$21 kpc between the pair occurred about 0.5 Gyr ago \citep[e.g. ]{brouillet91}, triggering a burst of star cluster formation at that time \citep{degrijs01}. The present starburst is associated with later nuclear infall of tidally disrupted interstellar material \citep[e.g. ][]{oconnell78}. 

If the two pairs of galaxies were similar, then we may expect to find tidal signatures in the ESO~565 pair in deep multi-wavelength imaging. But unlike M82 which is dominated by its disk, ESO~565--G019 is bulge-dominated. This may well render it more resilient to interaction-induced morphological changes (see, e.g., the simulations by \citealt{mihos94a}). Conversely, one may also expect AGN activity in M82, similar to that seen in ESO~565--G019. But despite long efforts to locate ongoing accretion onto any super-massive black hole, none has been identified so far \citep{g11_m82}, which perhaps emphasizes importance of supernova feedback in disrupting continuous accretion onto any AGN in M82 over the past 0.5 Gyr. 

Finally, we also mention the study of \citet{yuan10}, who investigate the co-evolution of starbursts and AGN in a large sample of IR-selected galaxy mergers. According to their optical spectral line diagnostics, \eso565\ is strongly AGN-dominated ($D_{\rm AGN}=0.8$ in their terminology). Overall, AGN dominance and total IR luminosity increase with the progression of a merger in their sample, but \eso565\ is unusual in being highly AGN-dominated despite its wide binary separation and moderate IR luminosity. This suggests that even if an interaction with ESO~565--G018 is ongoing, additional influences ought to be examined. 

\subsubsection{Past mergers}

The early-type morphological classification of \eso565\ suggests that it may already be in the final stages of a past major merger, with a mature spheroid being the end product and the remnants of the merger now providing the gas which obscures and feeds the nucleus. In this scenario, however, \eso565\ would not follow the now-popular merger and feedback-driven evolutionary scheme of symbiotic AGN and galaxy growth \citep[e.g. ][]{hopkins06, sanders88}, in which energetic feedback from accretion can clear away accreting gas and results in a \lq dead\rq\ super-massive black hole at the nucleus. The nucleus of \eso565\ is obviously surrounded by plentiful gas and growing at a radiatively-efficient rate at present. 

Instead, perhaps a more likely possibility is that a recent minor merger of a small system with ESO~565--G019 may have triggered enhanced nuclear activity, especially if the merger was \lq wet\rq\ and brought in a large amount of gas. Such mergers should be quite common, a hypothesis supported by simulations and observations, though they are not easy to identify. For instance, in simulations carried out by \citet{mihos94b}, it was found that a typical 10:1 merger of a disk system with a dwarf galaxy can briefly elevate SFRs by factors of a few, as appears to be the case for ESO~565--G019. With regard to observational evidence, \citet{fabbiano11} recently discovered a highly compact AGN pair in the spiral galaxy NGC~3393 -- thought to be the result of a minor merger in which both AGN of the pair are hidden by CT obscuration. Simulations by \citet{callegari11} have shown that in minor mergers, the smaller system of the pair undergoes distinct episodes of enhanced accretion that greatly promote its growth as compared to the growth of the larger black hole. This implies that many more such minor merger pairs with both black hole components actively accreting ought to be widely detectable. 

Fig.~\ref{fig:esor} shows an apparent asymmetry in the nuclear regions of \eso565, which could be disruption related to minor merger activity. If so, this would be the first minor-merger--driven reflection-dominated AGN in an early-type galaxy. A double AGN may then also be present. High angular resolution optical, X-ray and IR imaging in the future should be able to help distinguish between the various possibilities discussed above and also shed light on the relevance of any secular processes involved in triggering the observed nuclear activity.

\section{Conclusions}

We have presented the first X-ray spectrum of \eso565\ below 10 keV, as well as broadband X-ray spectral modeling. A reflection-dominated nuclear continuum and prodigious Fe fluorescence emission have been revealed with \suzaku. The line-of-sight obscuring column density is greater than 1.6$\times$10$^{24}$ cm$^{-2}$ (90\% confidence) and the Fe K$\alpha$ equivalent width with respect to the observed continuum is close to 1 keV. There is evidence of hard X-ray variability when comparing our \suzaku\ data with the long-term \swift/BAT average spectrum. This object is one of the few bona fide reflection-dominated AGN known to be hosted in an early-type galaxy. Mid- and far-infrared detections point to the likely presence of massive quantities of large-scale dust and on-going star-formation in the host. The most likely scenario for triggering nuclear activity is a minor merger, though there may also be an ongoing interaction with its neighbor ESO~565--G018. Our work shows that other reflection-dominated AGN will be uncovered in the nearby universe once the sensitivity of hard X-ray selection increases. Whether or not this population will turn out to be cosmologically significant will soon be investigated with the new generation of hard X-ray missions \nustar\ and \astroh.

\acknowledgments

PG acknowledges support from JAXA and STFC. The authors would like to express their thanks to \suzaku\ team members. \swift/BAT transient monitor data are used herein. This research has made use of the NASA/IPAC Extragalactic Database (NED) which is operated by the Jet Propulsion Laboratory, California Institute of Technology, under contract with the National Aeronautics and Space Administration. \wise\ is a project of Univ. California, Los Angeles, and Jet Propulsion Laboratory (JPL)/California Institute of Technology (Caltech), funded by the NASA. 2MASS is a project of the Univ. Massachusetts and the Infrared Processing and Analysis Center/Caltech, funded by NASA and the National Science Foundation. \akari\ is a project of the Japan Aerospace Exploration Agency with the participation of the European Space Agency. The NASA/IPAC Infrared Science Archive (IRSA) operated by JPL under contract with NASA, was used for querying the infrared databases. The HyperLEDA database was also useful in this work. The reviewer provided constructive criticisms which improved the paper. PG thanks S.F. H\"{o}nig for comments. KV would like to thank Y. Wadadekar and S. Barway for enlightening discussions.

\bibliographystyle{hapj}

\label{lastpage}
\end{document}